\documentclass[3p,times]{elsarticle}

\usepackage{lineno,hyperref}
\modulolinenumbers[5]

\journal{Physica A}

\usepackage{graphicx}
\usepackage{bm}
\usepackage{verbatim}
\usepackage{color}
\usepackage{amsmath,amssymb}

\newcommand{\be}{\begin{equation}}
\newcommand{\ee}{\end{equation}}
\newcommand{\ba}{\begin{eqnarray}}
\newcommand{\ea}{\end{eqnarray}}

\newcommand{\assumesErgodicity}{\refstepcounter{equation}\tag{{\theequation}*}}

\begin{document}

\begin{frontmatter}

\title{Fluctuation relations and strong inequalities for thermally isolated systems}

\author{Christopher Jarzynski}
\address{University of Maryland, College Park, MD 20742 USA}

\begin{abstract}
For processes during which a macroscopic system exchanges no heat with its surroundings, the second law of thermodynamics places two lower bounds on the amount of work performed on the system:
a weak bound, expressed in terms of a fixed-temperature free energy difference, $W \ge \Delta F_T$, and a strong bound, given by a fixed-entropy internal energy difference, $W \ge \Delta E_S$.
It is known that statistical inequalities related to the weak bound can be obtained from the nonequilibrium work relation, $\langle e^{-\beta W}\rangle = e^{-\beta\Delta F_T}$.
Here we derive an integral fluctuation relation $\langle e^{-\beta X}\rangle = 1$ that is constructed specifically for adiabatic processes, and we use this result to obtain inequalities related to the strong bound, $W \ge \Delta E_S$.
We provide both classical and quantum derivations of these results.
\end{abstract}


\end{frontmatter}


\centerline{\it This paper is dedicated to the memory of Christian Van den Broeck.}

\section{Introduction}
\label{sec:intro}

When a macroscopic system begins in equilibrium and then evolves in thermal isolation as a parameter $\lambda$ is varied at an arbitrary rate from $A$ to $B$, the process is said to be {\it adiabatic}.\footnote{
It is lamentable that the term {\it adiabatic} carries one meaning in thermodynamics, namely ``without exchange of heat'', and an entirely different meaning in dynamics and quantum mechanics, where it signifies ``infinitely slowly''.
Throughout this paper {\it adiabatic} will be used in its thermodynamic sense, and {\it quasi-static} will be used to indicate an infinitely slow process.
}
If instead the system exchanges energy with a thermal reservoir as the parameter is varied, then the process is {\it isothermal}.
In either case, the second law of thermodynamics places a lower bound on the work performed on the system:
for adiabatic processes we have
\be
\label{eq:2LS}
W_{adia} \ge \Delta E_S \equiv E(B,S) - E(A,S) = W_{adia}^{rev}
\ee
and for isothermal process,
\be
\label{eq:2LT}
W_{isoth} \ge \Delta F_T \equiv F[B,T] - F[A,T] = W_{isoth}^{rev}
\ee
Here, $E(\lambda,S)$ denotes the internal energy of an equilibrium state specified by the parameter value $\lambda$ and entropy $S$, and $F[\lambda,T]$ denotes the Helmholtz free energy of an equilibrium state identified by temperature rather than entropy.
Note that $(A,S)=[A,T]$ is the common initial equilibrium state for the two processes, and in Eq.~\ref{eq:2LT} the reservoir temperature is $T$.
Since an adiabatic process can be considered as a limiting case of an isothermal process, in which the energy exchanged as heat between system and reservoir is negligible, the bound in Eq.~\ref{eq:2LT} applies equally well to adiabatic processes:
\be
\label{eq:ineqChain}
W_{adia} \ge \Delta E_S \ge \Delta F_T
\ee
where the inequality $\Delta E_S \ge \Delta F_T$ can be established independently, as shown in the Appendix.

Eqs.~\ref{eq:2LS} - \ref{eq:ineqChain} follow from fundamental postulates of thermodynamics (see Appendix)~\cite{Callen1985,Finn1993}.
It is an important task of statistical mechanics to clarify how these results relate to the underlying microscopic dynamics of systems and reservoirs.
Fluctuation relations have emerged as a route both for deriving inequalities related to the second law, and for exploring how the second law applies to microscopic systems \cite{Evans2002,Bustamante2005,Cleuren2007,Esposito2009,Sekimoto2010,Kurchan2010,Campisi2011,Jarzynski2011,Seifert2012}.
In particular, the non-equilibrium work relation \cite{Jarzynski1997a,Jarzynski1997b}
\begin{subequations}
\label{eq:nwr-combined}
\be
\label{eq:nwr}
\left\langle e^{-\beta W}\right\rangle = e^{-\beta\Delta F_T}
\ee
which is valid for both adiabatic and isothermal processes (but is discussed predominantly in the latter context), rigorously implies the inequalities
\ba
\label{eq:2LT-av}
&\langle W\rangle \ge \Delta F_T \\
\label{eq:2LT-bound}
&{\rm Prob}(W \le \Delta F_T - \epsilon) \le e^{-\beta\epsilon}
\ea
\end{subequations}
Here and below, angular brackets $\langle\cdot\rangle$ indicate an ensemble average over realizations (repetitions) of the process in question, with initial conditions sampled from equilibrium, and $\beta^{-1} = kT$, where $k$ is Boltzmann's constant.
The left side of Eq.~\ref{eq:2LT-bound} denotes the probability of observing a value of work no greater than $\Delta F_T - \epsilon$, for arbitrary $\epsilon\ge 0$.
Thus Eq.~\ref{eq:nwr} both implies that the inequality $W \ge \Delta F_T$ is satisfied on average (Eq.~\ref{eq:2LT-av}), and places a strict bound on the probability of observing sizeable ``violations'' (Eq.~\ref{eq:2LT-bound}).

For adiabatic processes, Eq.~\ref{eq:nwr-combined} relates to the weak bound ($W_{adia} \ge \Delta F_T$) that appears in Eq.~\ref{eq:ineqChain}, but the strong bound ($W_{adia} \ge \Delta E_S$) has not been explored systematically in the context of fluctuation relations.
In the present paper we derive an integral fluctuation relation analogous to Eq.~\ref{eq:nwr}, but constructed with adiabatic processes in mind.
We then obtain corresponding analogues of Eqs.~\ref{eq:2LT-av} and \ref{eq:2LT-bound} and explore how these inequalities relate to the strong bound $W \ge \Delta E_S$.
Here and throughout the rest of the paper, we drop the subscript on $W_{adia}$, as we will concern ourselves only with adiabatic processes.

A monatomic ideal gas undergoing adiabatic compression or expansion provides a useful illustration of the results we will derive.
For this example, Eq.~\ref{eq:ineqChain} becomes
\be
\label{eq:ineqChain-gas}
W \ge \frac{3}{2} NkT \left[ \left( \frac{V_A}{V_B} \right)^{2/3} - 1 \right] \ge NkT \ln \left( \frac{V_A}{V_B} \right)
\ee
where the system volume $V$ plays the role of the parameter $\lambda$, and $V_A$ and $V_B$ are initial and final volumes.
The expressions appearing in Eq.~\ref{eq:ineqChain-gas} are familiar ones for the reversible adiabatic and isothermal work performed on an ideal gas \cite{Engel2006}.
For this example, the central results derived below (Eq.~\ref{eq:main}) can be rewritten as
\begin{subequations}
\label{eq:gasResults}
\ba
\label{eq:identity-gas}
&\left\langle e^{-\beta[\alpha E_f - E_i]}\right\rangle = 1 \\
\label{eq:2LS-av-gas}
&\langle W\rangle \ge \langle \Delta E_S \rangle \\
\label{eq:2LS-bound-gas}
&{\rm Prob}(W \le \Delta E_S - \epsilon) \le e^{-\beta^*\epsilon}
\ea
\end{subequations}
where $E_i$ and $E_f$ are the initial and final energies during one realization (hence $W=E_f-E_i$),
\be
\label{eq:alphadef}
\alpha = \left(\frac{V_B}{V_A}\right)^{2/3}
\quad,\quad
\beta^* = \alpha\beta
\ee
and
\be
\label{eq:DeltaES-gas}
\Delta E_S = (\alpha^{-1}-1) E_i = W_{adia}^{rev}
\ee
is the work that would be performed if the process were carried out adiabatically and reversibly.
Note that Eqs.~\ref{eq:2LS-av-gas} and \ref{eq:2LS-bound-gas} are related to the strong bound in Eq.~\ref{eq:ineqChain-gas}, $W\ge\Delta E_S$, whereas Eqs.~\ref{eq:2LT-av} and \ref{eq:2LT-bound} reflect only the weak bound, $W\ge\Delta F_T$.
Thus, while both Eqs.~\ref{eq:nwr} and \ref{eq:identity-gas} are valid integral fluctuation relations for the adiabatic compression or expansion of an ideal gas, the latter leads to stronger bounds on the work (Eqs.~\ref{eq:2LS-av-gas}, \ref{eq:2LS-bound-gas}) than the former Eqs.~\ref{eq:2LT-av}, \ref{eq:2LT-bound}).

In Sec.~\ref{sec:deriv} we derive our central results, Eqs.~\ref{eq:main} (a-c), within a classical, Hamiltonian model.
These results are identically valid regardless of the size of the system.
We then argue in Sec.~\ref{sec:2ndlaw} that for macroscopic systems, these results provide a derivation of the inequality $W \ge \Delta E_S$ -- to be more precise, they stringently constrain the probability distribution of observing violations of this inequality.
We sketch the quantum version of these results in Sec.~\ref{sec:quantum}, and end with a discussion in Sec.~\ref{sec:disc}.
The presentation is largely self-contained, with a few technical details relegated to the Appendix.

\section{Derivation of central results}
\label{sec:deriv}

Consider a classical system with $N$ degrees of freedom, described by a Hamiltonian $H(z,\lambda)$ or $H_\lambda(z)$, where $z$ denotes a point in $2N$-dimensional phase space.
For this parameter-dependent Hamiltonian we introduce the functions
\ba
\label{eq:Omegadef}
\Omega_\lambda(E) &=& \int dz \, \theta\left[ E - H_\lambda(z) \right] \\
\label{eq:Sigmadef}
\Sigma_\lambda(E)&=& \int dz \, \delta \left[ E - H_\lambda(z) \right] = \frac{\partial\Omega_\lambda}{\partial E} > 0
\ea
where $\theta(\cdot)$ is the unit step function, and the integrals -- which are assumed to converge -- are over phase space.
$\Omega_\lambda(E)$ is the volume of phase space enclosed by the energy shell $E$ of $H_\lambda$, and $\Sigma_\lambda(E)$ is akin to a surface area associated with this shell.
We use the term {\it energy shell} to denote the set of phase space points satisfying $H_\lambda(z)=E$.

Let us imagine that this system is prepared in equilibrium at temperature $T$ and parameter value $\lambda=A$, and then it evolves in thermal isolation as the parameter is varied from $\lambda_0=A$ to $\lambda_\tau=B$ according to a pre-determined protocol $\lambda_t$, with $0\le t\le\tau$.
For a given realization of the process, the initial microstate $z_0$ is sampled from the canonical distribution,
\be
\label{eq:eqA}
\pi_A^{c}(z_0) = 
\frac{1}{Z_A} e^{-\beta H_A(z_0)}
\ee
where $Z_A$ is the classical partition function.
We can imagine that prior to $t=0$ the system was allowed to equilibrate in weak contact with a thermal reservoir, and then disconnected from the reservoir.
During the interval $t\in[0,\tau]$, the system is described by a Hamiltonian phase space trajectory $z_t$ evolving under $H(z,\lambda_t)$.

By the first law of thermodynamics, the work performed on the system during this adiabatic process is given by
\be
\label{eq:Wdef}
W = E_f - E_i 
\ee
where
\be
\label{eq:E_if}
E_i = H_A(z_0) \quad,\quad E_f = H_B(z_\tau)
\ee
are initial and final energies along the trajectory.
Since the Hamiltonian evolution of the system is deterministic, both the final conditions $z_\tau$ and the work $W$ are functions of the initial conditions $z_0$.
An ensemble of realizations of the process, with $z_0$ sampled from equilibrium, generates a distribution of work values, reflecting the fluctuations inherent to a microscopic treatment.
When the system is macroscopic we expect this distribution to be exceedingly sharply peaked around its mean, thereby recovering the thermodynamic picture in which fluctuations are neglected.

In the adiabatic setting just described, Eq.~\ref{eq:nwr} is an identity and Eqs.~\ref{eq:2LT-av} and \ref{eq:2LT-bound} follow straightforwardly~\cite{Jarzynski1997a,Jarzynski1999a,Jarzynski2011}.
For the same adiabatic process and ensemble of realizations, we will now derive the results:
\begin{subequations}
\label{eq:main}
\ba
\label{eq:identity}
& \left\langle e^{-\beta X}\right\rangle = 1 \\
\label{eq:ineq}
& \left\langle X \right\rangle \ge 0 \\
\label{eq:bound}
& {\rm Prob}(X \le - \epsilon) \le e^{-\beta\epsilon}
\ea
\end{subequations}
with $X$ and a companion quantity, $Y$, defined as follows.
First, we define a function $E_A(\cdot)$ and its inverse $E_B(\cdot)$, through the equations
\begin{subequations}
\label{eq:EAEBdef}
\ba
\label{eq:EAdef}
\Omega_A(E_A(E)) &=& \Omega_B(E) \\
\label{eq:EBdef}
\Omega_A(E) &=& \Omega_B(E_B(E))
\ea
\end{subequations}
Then, for a realization $\{ z_t ; 0 \le t \le\tau \}$ we define
\begin{subequations}
\label{eq:XYdef}
\ba
\label{eq:Xdef}
X &=& E_A(E_f) - E_i \\
\label{eq:Ydef}
Y &=& E_f - E_B(E_i)
\ea
\end{subequations}
with initial and final energies $E_i$ and $E_f$ given by Eq.~\ref{eq:E_if}.

\begin{figure}[tbp]
\includegraphics[trim = 0in 2in 0in 1.4in , scale=0.5,angle=0]{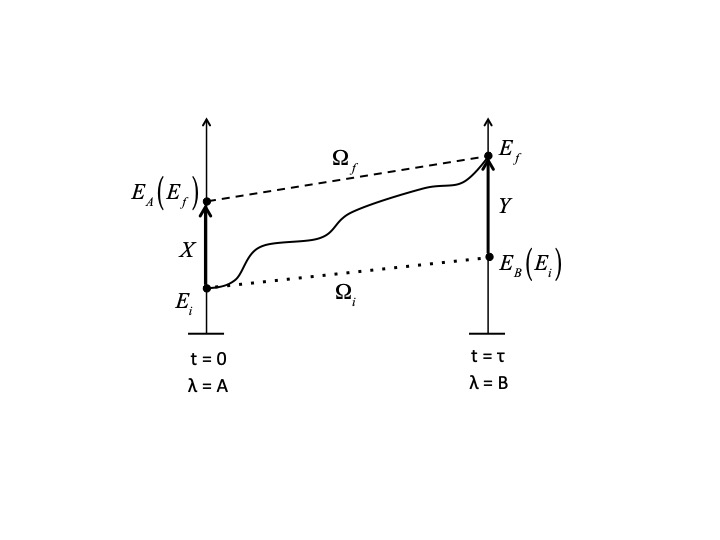}
\caption{
Vertical axes denote energies.
The curved solid line represents the energy of a trajectory $z_t$ evolving under $H(z,\lambda_t)$ (see Eq.~\ref{eq:E_if}).
The dotted and dashed lines connect states that share the same value of $\Omega_\lambda$ (Eqs.~\ref{eq:EAEBdef}, \ref{eq:Omega_if}).
The bold arrows depict $X$ and $Y$ (Eq.~\ref{eq:XYdef}), which are positive since $\Omega_f>\Omega_i$ for this trajectory (Eq.~\ref{eq:signs}).
For a trajectory with final energy $E_f < E_B(E_i)$ (not shown), we would have $\Omega_f<\Omega_i$ and $X,Y<0$.
}
\label{fig:schematic}
\end{figure}

As illustrated in Fig.~\ref{fig:schematic}, the function $E_A(\cdot)$ takes as input an energy shell $E$ of $H_B$, and it returns as output the energy shell of $H_A$ that encloses the same volume of phase space.
$E_B(\cdot)$ is defined conversely.
The chain rule applied to Eq.~\ref{eq:EAdef} gives us
\be
\label{eq:chain}
\Sigma_A(E_A(E)) \frac{dE_A}{dE}(E) = \Sigma_B(E)
\ee
and a similar result holds for Eq.~\ref{eq:EBdef}.
These identities will be used in Eqs.~\ref{eq:mainDeriv} and \ref{eq:dEdE} below.
Additionally, for any function $\phi(E)$, the identity
\be
\label{eq:usefulIdentity}
\int dz \, \phi(H_\lambda(z)) = \int dE \, \Sigma_\lambda(E) \, \phi(E)
\ee
follows by inserting $1 = \int dE \, \delta( E - H_\lambda)$ into the integral on the left, and then using Eq.~\ref{eq:Sigmadef}.

If we let
\be
\begin{aligned}
\label{eq:Omega_if}
\Omega_i &= \Omega_A(E_i) = \Omega_B(E_B(E_i)) \\
\Omega_f &= \Omega_B(E_f) = \Omega_A(E_A(E_f))
\end{aligned}
\ee
denote the initial and final values of $\Omega_\lambda$ along the trajectory, then Eq.~\ref{eq:XYdef} implies 
\be
\label{eq:signs}
{\rm sign}(X) = {\rm sign}(Y) = {\rm sign}(\Omega_f-\Omega_i)
\ee
as illustrated in Fig.~\ref{fig:schematic}.
Moreover, the definitions introduced above give us
\be
\label{eq:YWdE}
Y = W - \Delta E_\Omega
\ee
where $\Delta E_\Omega = E_B(E_i) - E_i$ is an energy change at fixed $\Omega$.

We now derive Eq.~\ref{eq:identity} by writing the average over realizations as an average over initial conditions $z_0$:
\ba
\label{eq:mainDeriv}
\left\langle e^{-\beta X} \right\rangle &=&
\int dz_0 \, \pi_A^{c}(z_0) \, e^{-\beta \left[ E_A(H_B(z_\tau)) - H_A(z_0) \right]} \nonumber \\
&=& \frac{1}{Z_A} \int dz_\tau \, e^{-\beta E_A(H_B(z_\tau))} \nonumber \\
&=& \frac{1}{Z_A} \int dE \, \Sigma_B(E) \, e^{-\beta E_A(E)} \nonumber \\
&=& \frac{1}{Z_A} \int dE \, \Sigma_A(E_A(E)) \,  \frac{dE_A}{dE}(E) e^{-\beta E_A(E)} \nonumber \\
&=& \frac{1}{Z_A} \int d E_A \, \Sigma_A(E_A) \, e^{-\beta E_A} \nonumber \\
&=& \frac{1}{Z_A} \int dz \, e^{-\beta H_A(z)} = 1
\ea
Here we first performed a change in the variables of integration from initial to final conditions $z_\tau=z_\tau(z_0)$; the associated Jacobian is unity, $\vert \partial z_\tau/\partial z_0\vert = 1$, by Liouville's theorem.
We then used Eqs.~\ref{eq:chain} (once) and \ref{eq:usefulIdentity} (twice) to get to the last line.
From Eq.~\ref{eq:identity} we obtain Eq.~\ref{eq:ineq} via Jensen's inequality, $\langle e^x\rangle \ge e^{\langle x\rangle}$, and Eq.~\ref{eq:bound} as follows:
\ba
{\rm Prob}(X \! \le \! -\epsilon) = \int_{-\infty}^{-\epsilon} dX \, \rho(X)
&=& \left\langle \theta(-\epsilon-X)\right\rangle \nonumber\\
&\le&  \left\langle \theta(-\epsilon-X) \, e^{\beta(-\epsilon-X)}\right\rangle \nonumber\\
&\le& e^{-\beta\epsilon} \left\langle e^{-\beta X}\right\rangle = e^{-\beta\epsilon}
\ea
where $\rho(X)$ is the statistical distribution of values of $X$, over the ensemble of realizations.

Eq.~\ref{eq:main} was obtained using minimal assumptions, namely canonically distributed initial conditions (Eq.~\ref{eq:eqA}) and Hamiltonian evolution for $t \in [0,\tau]$.
If we additionally assume the system's Hamiltonian dynamics are {\it ergodic}~\cite{Dorfman1999} for all fixed values of $\lambda$, then $\Omega_\lambda$ remains invariant when $\lambda$ is changed quasi-statically~\cite{Hertz1910a,Hertz1910b,Anosov1960,Kasuga1961a,Kasuga1961b,Kasuga1961c,Ott1979,Lochak1988}:
\be
\assumesErgodicity
\label{eq:qs}
\Omega_B(E_f) \stackrel{qs}{=} \Omega_A(E_i)
\ee
Here $qs$ denotes the quasi-static limit ($\tau\rightarrow\infty$) of infinitely slow variation from $\lambda_0=A$ to $\lambda_\tau=B$, and the asterisk by the equation number specifies that the property of ergodicity was assumed to obtain this result.
We then have
\be
\assumesErgodicity
\label{eq:Xqs}
X \stackrel{qs}{=} 0
\ee
for every trajectory, equivalently
\be
\assumesErgodicity
\label{eq:rhoXqs}
\rho(X) \stackrel{qs}{=} \delta (X)
\ee
By Eq.~\ref{eq:signs}, we also get $Y\stackrel{qs}{=}0$ (*) for every trajectory.
However, the value of work, $W$, generally differs from one realization to the next.

If the assumption of ergodicity is not satisfied, then even for quasi-static driving the value of $X$ generically depends on the initial conditions $z_0$: for some realizations we will have $X>0$ and for others $X<0$.
In this situation we have $\langle X\rangle > 0$, since Eq.~\ref{eq:identity} implies that $\langle X\rangle = 0$ if and only if $X=0$ for every realization.
We emphasize that while ergodicity was assumed to obtain Eqs.~\ref{eq:qs} - \ref{eq:rhoXqs}, it was not required in the derivation of Eq.~\ref{eq:main}.

Let us consider these results in the context of the ideal gas example introduced earlier.
For this example we have
\be
\label{eq:OmegaGas}
\Omega_V(E) = V^N \cdot \frac{(2\pi mE)^\nu}{\Gamma(\nu+1)}
\ee
where $\nu = 3N/2$ and $\Gamma(\cdot)$ is the Gamma function.
The first factor on the right accounts for the arrangement of $N$ particles in a box of volume $V$, and the second factor is the volume of momentum space enclosed by a $3N$-dimensional hypersphere of radius $\sqrt{2mE}$~\cite{Gibbs1902p93}.
From Eq.~\ref{eq:OmegaGas} it follows that
\be
E_A(E) = \alpha E \quad,\quad E_B(E) = \frac{E}{\alpha}
\ee
with $\alpha = (V_B/V_A)^{2/3}$ as in Eq.~\ref{eq:alphadef}, and therefore (see Eqs.~\ref{eq:DeltaES-gas}, \ref{eq:XYdef} and \ref{eq:YWdE})
\be
\label{eq:XYgas}
X = \alpha E_f - E_i = \alpha Y = \alpha (W - \Delta E_S) \quad .
\ee
Eq.~\ref{eq:main} then becomes Eq.~\ref{eq:gasResults}.
Thus for this example we have shown both that the strong bound $W\ge\Delta E_S$ is satisfied on average, Eq.~\ref{eq:2LS-av-gas}, and that violations of this bound are exponentially suppressed, Eq.~\ref{eq:2LS-bound-gas}.
This example also illustrates the comments following Eq.~\ref{eq:rhoXqs}: in the quasi-static limit $X=Y=0$ for every realization, but the work distribution has a finite width, as shown by explicit calculation in Ref.~\cite{Crooks2007}.
For non-quasi-static compression or expansion, Bena, Van den Broeck and Kawai~\cite{Bena2005} have analyzed the work distribution for an exactly solvable model of an ideal gas (a {\it Jepsen gas}); their approach may provide another tractable model for studying the results of the present paper.

Note that while the result ${\rm sign}(X) = {\rm sign}(Y)$ (Eq.~\ref{eq:signs}) is valid by construction, the linear relation $X=\alpha Y$ (Eq.~\ref{eq:XYgas}) is a non-generic  feature of the ideal gas.
It is this linearity that allows us to obtain the inequality $\langle W\rangle \ge\langle\Delta E_S\rangle$ from the equality $\langle e^{-\beta X}\rangle = 1$, for this particular example.

\section{Connection to the inequality $W\ge\Delta E_S$ for macroscopic systems}
\label{sec:2ndlaw}

Eqs.~\ref{eq:main} (a-c) are identically valid within our Hamiltonian setup.
To connect these results to the strong bound $W\ge\Delta E_S$ we now explicitly assume a macroscopic system of interest, with the proverbial $N\sim 10^{23}$ degrees of freedom.

Next, we need a prescription for assigning a value of entropy $S$ to any energy shell $(E,\lambda)$, with monotonic dependence $\partial S/\partial E > 0$.
Once this assignment is made, we can invert the function $S(E,\lambda)$ to get $E(\lambda,S)$, which is needed to define $\Delta E_S$  (Eq.~\ref{eq:2LS}).
We choose to define entropy as follows:
\be
\label{eq:Sdef}
S(E,\lambda) = k \ln \frac{\Omega_\lambda(E)}{h^{3N} \prod_s N_s!}
\ee
where $h$ is Planck's constant, and the product in the denominator is over the various chemical species that comprise the system, with $\sum_s N_s = N$.
This definition combines with Eqs.~\ref{eq:2LS} and \ref{eq:EBdef} to give
\be
\label{eq:DEs}
\Delta E_S = \Delta E_\Omega = E_B(E_i) - E_i
\ee

Eq.~\ref{eq:Sdef} is the Gibbs (or ``volume'') entropy.
An alternative definition is the Boltzmann (``area'') entropy, with $\Sigma_\lambda$ appearing in place of $\Omega_\lambda$  in Eq.~\ref{eq:Sdef}.\footnote{
The Boltzmann entropy additionally requires an arbitrary factor with units of energy, $\delta E$, to make the argument of the logarithm dimensionless.
}
Yet another definition is the canonical entropy, that is the Shannon entropy of the canonical distribution whose average energy is equal to $E$.
The interested reader will find a lively debate on the relative merits and demerits of these microscopic definitions of entropy in Refs.~\cite{Dunkel2014,Sokolov2014,Vilar2014,Frenkel2015,Dunkel2014a,Schneider2014,Dunkel2014b,Hilbert2014,Campisi2015,Swendsen2016};
see also Ref.~\cite{Lebowitz1993} for a lucid discussion of Boltzmann's entropy in the context of time's arrow.
For present purposes, the salient feature of all three definitions is that they converge to the Clausius (thermodynamic) entropy in the macroscopic limit, which is the limit considered in this section.

We further assume that the system dynamics are, for practical purposes, ergodic: when the system evolves under the Hamiltonian $H_\lambda(z)$ at fixed $\lambda$, from arbitrary initial conditions $z_0$, then after a characteristic relaxation time $t_{relax}$ its microstate may be viewed as a random sample from the microcanonical distribution
\be
\pi_\lambda^{\mu c}(z) = \frac{1}{\Sigma_\lambda} \delta(E-H_\lambda)
\ee
The assumption of ergodicity reflects, at the microscopic level, the macroscopic property of self-equilibration: if a thermodynamic system is prepared in an arbitrary initial state and then allowed to evolved undisturbed, at fixed $\lambda$ and constant energy $E$, then it relaxes spontaneously to an equilibrium state whose properties are determined uniquely by the values $(E,\lambda)$.
The temperature $T(E,\lambda)$ of this state is given by
\be
\label{eq:T}
\frac{1}{T} =  \frac{\partial S}{\partial E} = \frac{\Sigma_\lambda(E)}{\Omega_\lambda(E)}
\ee
We defer a further discussion of the assumption of ergodicity to Sec.~\ref{sec:disc}.

We are now in a position to investigate how Eq.~\ref{eq:main} relates to the macroscopic bound on adiabatic work, $W \ge \Delta E_S$.
For a quasi-static process the assumption of ergodicity implies Eq.~\ref{eq:qs}, which combines with our definition of entropy to give
\be
\label{eq:Sqs}
S_f \stackrel{qs}{=} S_i
\ee
where $S_i=S(E_i,A)$ and $S_f=S(E_f,B)$.
We further have
\be
\label{eq:Wqs}
W \stackrel{qs}{=} E_B(E_i)-E_i = \Delta E_S
\ee
using Eq.~\ref{eq:DEs}.
These results agree with macroscopic expectations that for quasi-static processes the system entropy remains constant, and the bound in Eq.~\ref{eq:2LS} is saturated.
The connection between the quasi-static invariance of  the Gibbs entropy and the second law of thermodynamics has been explored by Berdichevsky~\cite{Berdichevsky1988,Berdichevsky1997} and Campisi~\cite{Campisi2005}.

For a process of arbitrary duration, Eq.~\ref{eq:YWdE} can now be written as
\be
\label{eq:YWES}
Y = W - \Delta E_S
\ee
As we have shown that $\langle X\rangle \ge 0$ (Eq.~\ref{eq:ineq}), it is tempting to conclude that $\langle Y\rangle \ge 0$ since ${\rm sign}(X)={\rm sign}(Y)$ (Eq.~\ref{eq:signs}).
However, the former inequality does not logically imply the latter -- unless (as with the ideal gas, Eq.~\ref{eq:XYgas}) $X$ is simply proportional to $Y$.
Let us investigate more carefully what Eqs.~\ref{eq:ineq} and \ref{eq:bound} reveal about the strong bound on work, $W\ge\Delta E_S$, beginning with qualitative observations.

Generically, for a macroscopic system the distributions of both $E_i$ and $E_f$ are exceedingly sharply peaked.
We expect the distribution of $X$ to be similarly sharply peaked, i.e.\ $\sigma_X \ll \left\vert \langle X\rangle \right\vert$, where $\sigma$ denotes standard deviation.
Eq.~\ref{eq:ineq} then gives us 
\be
\label{eq:sharplyPeaked}
\sigma_X \ll \langle X \rangle
\ee
which implies that $X > 0$, and therefore $Y > 0$ (by Eq.~\ref{eq:signs}), for nearly every realization of the process.
In this way, based on generic expectations that fluctuations are negligible in the thermodynamic limit, we loosely conclude that the strong lower bound on adiabatic work is satisfied ``almost always''.
These expectations also combine with Eq.~\ref{eq:ineq} to suggest strongly that the inequality $W\ge\Delta E_S$ is satisfied on average, for macroscopic systems.
It is difficult, however, to translate generic expectations into reliable derivations of inequalities such as Eq.~\ref{eq:sharplyPeaked}, particularly for irreversible processes.

For a quantitative treatment we turn to Eq.~\ref{eq:bound}, which places a rigorous bound on values of $X$ in the ``thermodynamically forbidden'' tail $X<0$ of the distribution $\rho(X)$.
We wish to translate this result into a similar bound on the probability to observe a work value $W \le \Delta E_S - \epsilon$, for a generic adiabatic process starting from the equilibrium state $(A,S)=[A,T]$.
To make the argument, let us choose a dimensionless number $n$ that satisfies
\be
1 \ll n \ll N\sim 10^{23}
\ee
The choice $n=1000$ will do just fine.
Let the notation $X\lesssim 0$ denote the window of $X$ values
\be
\label{eq:window}
-nkT \le X \le 0
\ee
where $X$ is negative, but not macroscopically so.
This window corresponds to modest violations of the inequality $W \le \Delta E_S$.
We will refer to the region $X\lesssim 0$ as the {\it near tail} of the distribution $\rho(X)$, and to the region $X < nkT$ as the {\it far tail} of the distribution.
As shown in the Appendix, in the near tail $Y$ is linear in $X$:
\be
\label{eq:linear}
Y = \frac{T^*}{T} X \equiv rX
\quad,\quad X\lesssim 0
\ee
aside from a negligible correction that scales like $N^{-1}$.
Here $T^*$ is the temperature of the macroscopic state $(B,S)$, that is, the final temperature that would be reached if the process were performed quasi-statically.
Eq.~\ref{eq:bound} can then be rewritten as
\be
{\rm Prob}(Y \le - r\epsilon) \le e^{-\beta\epsilon}
\ee
Replacing $\epsilon$ by $\epsilon/r$ and using Eq.~\ref{eq:YWES}, we arrive at
\be
\label{eq:workBound}
{\rm Prob}(W \le \Delta E_S - \epsilon) \le e^{-\beta^*\epsilon}
\quad,\quad
\frac{1}{\beta^*} = kT^*
\ee
This result places a tight bound on the probability to observe sizeable violations of the strong inequality $W \ge \Delta E_S$ when a macroscopic system undergoes an adiabatic process.
For instance it implies that the probability to observe a work value that undershoots $\Delta E_S$ by at least $100\ kT^*$ is no greater than $e^{-100}$, which is fantastically small -- even though a work value $W = \Delta E_S - 100\ kT^*$ represents only a tiny violation (on the macroscopic scale) of the second law.
Eq.~\ref{eq:workBound} is valid in the near tail of the work distribution, that is in the region of validity of the linear approximation in Eq.~\ref{eq:linear}.
Thus it might not accurately apply to the far tail ($X < -nkT$), but in that region Eq.~\ref{eq:bound} implies that the probability becomes entirely negligible.

The monatomic ideal gas illustrates the results discussed in this section.
Combining Eqs.~\ref{eq:OmegaGas} and \ref{eq:Sdef} with Stirling's approximation, $\ln[\Gamma(\nu+1)] = \ln \nu ! \approx \nu\ln\nu - \nu$, we obtain
\be
\label{eq:sackur}
S(E,V) = kN \ln \left[ \frac{V}{N} \left( \frac{4\pi mE}{3Nh^2} \right)^{3/2} \right] + \frac{5}{2} kN
\ee
This is the Sackur-Tetrode equation for the entropy of an ideal gas, illustrating that the Gibbs entropy agrees with the thermodynamic entropy for macroscopic systems.
Using Eq.~\ref{eq:T} to obtain the familiar result $E = (3/2)NkT$, it follows from Eq.~\ref{eq:sackur} that when the gas is compressed or expanded adiabatically and quasi-statically (i.e.\ at constant entropy), the quantity $TV^{2/3}$ remains constant.
Thus for the temperature ratio in Eq.~\ref{eq:linear} we get
\be
\frac{T^*}{T} = \left( \frac{V_A}{V_B} \right)^{2/3} = \frac{1}{\alpha}
\ee
This result provides a modest consistency check, as it verifies that the quantity $\beta^*$ appearing in Eq.~\ref{eq:workBound} -- which was derived with a generic macroscopic system in mind -- is the same as the quantity $\beta^*$ appearing in Eq.~\ref{eq:2LS-bound-gas}, which was obtained by analyzing Eq.~\ref{eq:bound} for the specific example of a monatomic ideal gas.

\section{Quantum analysis}
\label{sec:quantum}

Now consider a quantum system described by a Hamiltonian operator $\hat H(\lambda)$, and for simplicity assume a non-degenerate energy spectrum for all values of the parameter $\lambda$.
Imagine a process that follows the two-point measurement procedure:
the system is prepared in equilibrium at $\lambda=A$ and temperature $T$, then the initial energy is measured, then the system evolves unitarily under $\hat H(\lambda_t)$ as the parameter is varied from $\lambda_0=A$ to $\lambda_\tau=B$, and finally another energy measurement is performed.
For a given realization, let $\vert m_A\rangle$ and $\vert n_B\rangle$ specify the initial and final energy eigenstates, with energies $E_m^A$ and $E_n^B$ -- these are the outcomes of the energy measurements.
The first measurement samples the initial energy eigenstate from a statistical mixture described by the density matrix
\be
\hat\pi_A^{c} = \frac{1}{Z_A} \sum_m \vert m_A \rangle \langle m_A\vert \, e^{-\beta E_m^A}
\quad,\quad
Z_A = \sum_m e^{-\beta E_m^A}
\ee
whereas the second measurement is projective, causing the pure state $\vert\psi(t=\tau)\rangle$ to ``collapse'' onto an energy eigenstate.
The subscripts in the notation $\vert m_A\rangle$ and $\vert n_B\rangle$ are labels indicating that these are eigenstates of $\hat H(A)$ and $\hat H(B)$.

For a given realization of the process just described, the measured values $E_m^A$ and $E_n^B$ are interpreted as the initial and final energies of the system, and the two-point definition of work~\cite{Kurchan2000,Tasaki2000,Mukamel2003,Talkner2007}
\be
\label{eq:qWdef}
W = E_n^B - E_m^A
\ee
is the quantum analogue of classical work, Eq.~\ref{eq:Wdef}.
Since the quantum number is invariant under quasi-static driving, the definitions
\ba
\Delta E_S &=& E_m^B - E_m^A \\
\label{eq:qXdef}
X &=& E_n^A - E_m^A \\
Y &=& E_n^B - E_m^B
\ea
provide natural quantum analogues of the corresponding classical quantities (Eqs.~\ref{eq:2LS}, \ref{eq:XYdef}).
These definitions satisfy
\be
Y= W - \Delta E_S \quad,\quad {\rm sign}(X) = {\rm sign}(Y)
\ee
(compare with Eqs.~\ref{eq:signs}, \ref{eq:YWES}).
In the quasi-static limit we have $W \stackrel{qs}{=} \Delta E_S$ and $X,Y \stackrel{qs}{=} 0$ for every realization.

The identity $\langle e^{-\beta W}\rangle = e^{-\beta\Delta F_T}$ is easily established from the properties of unitarity~\cite{Kurchan2000,Tasaki2000,Mukamel2003}.
The result $\langle e^{-\beta X}\rangle = 1$ is equally easily derived:
\be
\label{eq:qderiv}
\begin{aligned}
\left\langle e^{-\beta X}\right\rangle &= \sum_{m,n} \left[ \frac{e^{-\beta E_m^A}}{Z_A} \left\vert U_{nm} \right\vert^2 \right] \, e^{-\beta X} \\
&= \frac{1}{Z_A} \sum_n e^{-\beta E_n^A} \sum_m \left\vert U_{nm} \right\vert^2  = 1
\end{aligned}
\ee
where $U_{nm} = \langle n_B \vert \hat U \vert m_A\rangle$ and $\hat U$ is the unitary time-evolution operator from $t=0$ to $t=\tau$.
In Eq.~\ref{eq:qderiv}, the quantity in square brackets is the joint probability for the two measurement outcomes to give states $\vert m_A\rangle$ and $\vert n_B\rangle$, and unitarity is invoked in the second line.
(Following Refs.~\cite{Kafri2012,Rastegin2013,Rastegin2014,Albash2013,Manzano2015,Smith2018}, Eq.~\ref{eq:qderiv} can readily be extended to {\it unital} dynamics.)
Eqs.~\ref{eq:ineq} and \ref{eq:bound} follow immediately, for the quantum case, from Eq.~\ref{eq:qderiv}, and the former gives us
\be
\langle E_n^A\rangle \ge \langle E_m^A\rangle
\ee
This result suggests (but does not rigorously imply) that the final quantum number is, on average, greater than the initial quantum number, i.e. $\langle n\rangle \ge \langle m\rangle$.

For a macroscopic quantum system undergoing an adiabatic process, arguments like those of Sec.~\ref{sec:2ndlaw} can be used to connect Eq.~\ref{eq:main}, in its quantum incarnation, to the inequality $W \ge \Delta E_S$.
We will not present those arguments here, primarily to avoid repeating the content of Sec.~\ref{sec:2ndlaw}, but also because the macroscopic quantum scenario seems excessively idealized.
It is hard to imagine an experimental situation in which a macroscopic quantum system is so utterly isolated from its environment that it evolves unitarily as an external parameter is varied from one value to another -- surely a wayward photon or gas molecule will scatter off the system, spoiling unitarity.
Of course, similar charges may be brought against the notion of a classical, macroscopic system evolving entirely under Hamiltonian dynamics.
These considerations highlight the idealizations that are made (and should always be kept in mind) when choosing specific dynamical equations of motion to model the evolution of a many-particle system.

\section{Discussion}
\label{sec:disc}

The problem addressed in this paper was motivated by the observation that for isothermal process the work relation Eq.~\ref{eq:nwr} can be used to establish both that the inequality $W_{isoth}\ge\Delta F_T$ is satisfied on average (Eq.~\ref{eq:2LT-av}), and that violations of this inequality are suppressed exponentially (Eq.~\ref{eq:2LT-bound}) -- these results hold independently of system size.
The original goal was to derive corresponding results for adiabatic processes.
This program was not entirely successful, as the inequality $\langle W_{adia}\rangle \ge \langle\Delta E_S\rangle$ has not been obtained, and the bound represented by Eq.~\ref{eq:bound} does not translate directly into a bound on the distribution of adiabatic work values.
(The ideal gas example provides a happy exception to these negative statements.)
Nevertheless, for macroscopic systems we have presented both qualitative and quantitative arguments connecting our central result, Eq.~\ref{eq:main}, to the inequality $W_{adia} \ge \Delta E_S$.
In particular, we have shown that the probability to observe violations of this inequality decays exponentially or faster in the near tail of the work distribution ($-nkT^*\le Y\le 0$), and is entirely negligible in the far tail ($Y < -nkT^*$).
This is consistent with empirical observations that macroscopic violations of the strong inequality are never observed.
It remains an open problem to determine whether these results can be strengthened.

We now briefly discuss a few issues related to the results of Secs.~\ref{sec:intro} - \ref{sec:quantum}, as well as avenues for future work.

In Sec.~\ref{sec:2ndlaw} it was assumed that the system's dynamics are ergodic ``for practical purposes'', at fixed $\lambda$.
Ergodicity has a precise mathematical meaning~\cite{Dorfman1999}, which in the present context can be paraphrased thus: a Hamiltonian trajectory wanders arbitrarily close to all points on the energy shell, in the infinite-time limit.
This property has been proven rigorously for only a limited number of Hamiltonian system, such as certain billiard systems~\cite{Sinai1970,Bunimovich1979,Wojtkowski1986}.
Moreover, it is recognized that a generic Hamiltonian has a (non-ergodic) mixed phase space consisting of islands of stable motion scattered across a chaotic sea~\cite{Lichtenberg1992}.
Despite these caveats, in classical statistical mechanics ergodicity is often taken as a working hypothesis for many-particle systems~\cite{Dorfman1999}.
In effect, this hypothesis assumes that the stable islands occupy a negligible fraction of phase space, and that away from these islands trajectories sail the chaotic sea ergodically.
This is the sense in which ergodicity was assumed ``for practical purposes'' in Sec.~\ref{sec:2ndlaw}.

If we let $t_{erg}$ denote a characteristic timescale required for a trajectory to ergodically explore all regions of the energy shell, then for a  macroscopic system this timescale is absurdly long, say $t_{erg} \sim e^{10^{23}}$.
A more relevant quantity is the {\it mixing time} over which a trajectory ``forgets'' where it has been (see Ref.~\cite{Dorfman1999} for a proper definition), which is comparable to the relaxation time required for a system to self-equilibrate, $t_{mix}\sim t_{relax} \lll t_{erg}$.
The closely related properties of ergodicity and mixing, when combined with notions of {\it typicality}~\cite{Lebowitz1999}, provide insight into the self-equilibration of macroscopic systems on accessible timescales.

For an adiabatic process of the sort considered in this paper, Eqs.~\ref{eq:nwr}, $\langle e^{-\beta W}\rangle = e^{-\beta\Delta F_T}$, and \ref{eq:identity}, $\langle e^{-\beta X}\rangle = 1$, generally represent two distinct predictions, both of which apply to the same process. (The former additionally applies to isothermal processes, while the latter does not.)
In the special case of a {\it cyclic} adiabatic process, for which $H_A(z)=H_B(z)$, the definitions used in this paper imply $\Delta E_S = \Delta E_\Omega = \Delta F_T=0$ and $W=X=Y$.
Thus for cyclic adiabatic processes, Eqs.~\ref{eq:nwr} and \ref{eq:identity} are equivalent.

For processes identical to the ones analyzed in Secs.~\ref{sec:deriv} and \ref{sec:quantum} of the present paper -- that is, canonically sampled initial conditions\footnote{
In fact, Tasaki and Campisi require only the weaker assumption that initial conditions be sampled from a distribution that is a monotonically non-increasing function of $H_A$.
}
followed by Hamiltonian or unitary evolution -- Tasaki~\cite{Tasaki2000a} and Campisi~\cite{Campisi2008} have derived inequalities that are expressed in our notation as follows (see Eq.\ 9 of Ref.~\cite{Tasaki2000a} and Eqs.\ 16 and 36 of Ref.~\cite{Campisi2008}):
\be
\label{eq:TasakiCampisi}
\langle \ln\Omega_f\rangle \ge \langle \ln\Omega_i\rangle
\quad,\quad
\left\langle \ln n \right\rangle \ge \left\langle \ln m \right\rangle
\quad,\quad
\left\langle \ln\left(n+\frac{1}{2}\right)\right\rangle \ge \left\langle \ln\left(m+\frac{1}{2}\right)\right\rangle
\ee
Thus in the classical case the Gibbs entropy does not decrease (on average).
In the quantum case the same statement remains true if we replace the argument of the logarithm in Eq.~\ref{eq:Sdef} by the quantum number $n$~\cite{Tasaki2000a}, or else by $n + (1/2)$~\cite{Campisi2008}.
It would be useful to clarify the relationship between these inequalities and the ones obtained in the present paper.
While Eq.~\ref{eq:TasakiCampisi} does not follow directly from our result $\langle e^{-\beta X}\rangle = 1$, perhaps it can be obtained from another, yet-to-be-derived integral fluctuation relation.
It would also be interesting to investigate whether the approaches taken in Refs.~\cite{Tasaki2000a,Campisi2008} can be used to obtain bounds, analogous to Eq.~\ref{eq:workBound}, on the probability distribution of violations of the strong inequality $W \ge \Delta E_S$, or of the related inequality $\Omega_f \ge \Omega_i$.
Note that while $N\ggg 1$ was assumed in deriving Eq.~\ref{eq:workBound} in Sec.~\ref{sec:2ndlaw}, this limit was not assumed by Tasaki and Campisi when deriving Eq.~\ref{eq:TasakiCampisi}.

While this paper has focused on the integral fluctuation relation $\langle e^{-\beta X}\rangle=1$, it would be interesting to derive a detailed counterpart, analogous to the Crooks fluctuation relation~\cite{Crooks1998,Crooks1999} involving conjugate ``forward'' and ``reverse'' processes.
It may also be interesting to explore how Eq.~\ref{eq:identity} is modified by measurement and feedback, and whether an analogous Sagawa-Ueda-like fluctuation relation~\cite{Sagawa2010} could be derived.
As a specific example, suppose that after sampling the initial microstate $z_0$ from the canonical distribution $\pi_A^c$, the energy $E_i = H_A(z_0)$ is measured, and a protocol for varying the parameter $\lambda$ is then selected on the basis of the measurement outcome.
Such scenarios have been considered for cyclic processes in Refs.~\cite{Parrondo2010,Vaikuntanathan2011}, where it was shown that the protocol could be chosen so as to extract work, both on average ($\langle W\rangle <0$) and for nearly every realization of the process.

Throughout the paper we have assumed that initial conditions are sampled canonically, but for a thermally isolated system one could also consider sampling from a microcanonical distribution.
Deriving fluctuation relations in this case is challenging due to the singular nature of the microcanonical distribution, but Adib~\cite{Adib2005} and Cleuren, Van den Broeck and Kawai~\cite{Cleuren2006a} have developed approaches for addressing this challenge.
It would be useful to investigate whether these approaches might lead to microcanonical versions of Eq.~\ref{eq:main} of the present paper.

It would also be interesting to demonstrate the integral fluctuation relation Eq.~\ref{eq:identity} in the laboratory.
To date, experiments related to classical fluctuation relations have focused mostly on isothermal rather than adiabatic processes~\cite{Ciliberto2017}, and it may prove challenging first to equilibrate a microscopic system with a thermal reservoir and then to maintain the system in thermal isolation while $\lambda$ is varied from $A$ to $B$.
However, one can envision an experiment involving a macroscopic system such as a pendulum, for which initial conditions are determined not through contact with a thermal reservoir, but rather are imposed ``by hand'', using a computer code to generate random samples from a canonical distribution at a desired effective temperature (or, equivalently, at a chosen effective value of Boltzmann's constant $k$).
If the system is then reasonably well isolated from sources of friction while a parameter (say, the pendulum length) is varied with time, then its evolution will be nearly Hamiltonian, and both Eqs.~\ref{eq:nwr} and \ref{eq:identity} can be demonstrated experimentally by generating sufficiently many trajectories of this macroscopic system.
This approach amounts to ``experiment by analog simulation'' -- note that the true microscopic degrees of freedom (e.g.\ the atoms that make up the pendulum) are ignored here, whereas one or a few collective degrees (such as the angle of oscillation of the pendulum) are treated as effective microscopic degrees of freedom.

In contrast with the classical case, recent experimental tests of quantum fluctuation relations have focused on thermally isolated systems~\cite{Batalhao2014,An2015,Batalhao2015,Cerisola2017,Medeiros2018,Nagilhoo2018}.
In particular, in Ref.~\cite{An2015} An {\it et al} used a trapped-ion setup to implement a process involving a particle in a forced harmonic potential.
For each realization of the process the initial energy eigenstate was drawn from a canonical distribution, and the final eigenstate was determined by projective measurement, producing a distribution of work values that was then confirmed to be in agreement with Eq.~\ref{eq:nwr}.
For the trapped-ion system of Ref.~\cite{An2015}, the eigenvalues of $\hat H_B$ are shifted from those of $\hat H_A$ by a constant value $E_n^B - E_n^A = \Delta F_T$, which implies $X = W - \Delta F_T$ (see Eqs.~\ref{eq:qWdef} and \ref{eq:qXdef}).
Thus Eqs.~\ref{eq:nwr} and \ref{eq:identity} are equivalent for this particular system.

\section*{Dedication}
This paper is dedicated to the memory of Christian Van den Broeck, a wonderful colleague who thought deeply and creatively, and who inspired many of us in the statistical physics community.

\section*{Acknowledgments}
The author gratefully acknowledges support from the U.S. National Science Foundation under grant DMR-1506969.

\section*{Appendix}

\noindent{\it Macroscopic derivation of the inequality chain $W_{adia} \ge \Delta E_S \ge \Delta F_T$.}

Consider a process in which a thermally isolated system starts in equilibrium state $(A,S)$, then a parameter $\lambda$ is varied from $A$ to $B$ over a finite interval of time, and subsequently the isolated system self-equilibrates at fixed $\lambda=B$, and at constant internal energy, ending in an equilibrium state $(B,S^\prime)$.
The notation
\be
(A,S) \xrightarrow[adia]{irrev} (B,S^\prime)
\ee
indicates that the process is both adiabatic and (in general) irreversible.
The first and second laws of thermodynamics give us
\be
W_{adia} = E(B,S^\prime) - E(A,S) \quad,\quad S^\prime\ge S
\ee
Since $\partial E/\partial S = T > 0$ we have $E(B,S^\prime) \ge E(B,S)$ hence $W_{adia} \ge E(B,S) - E(A,S) = \Delta E_S$.
Note that no work is performed during the relaxation stage, at fixed $\lambda=B$.

To establish the inequality
\be
\label{eq:strongIneq}
\Delta E_S \ge \Delta F_T
\ee
imagine that the system undergoes a three-stage cyclic process, beginning and ending in the state $(A,S) = [A,T]$: \\
(i) The system is driven {\it adiabatically and reversibly} from $(A,S)$ to $(B,S)$ by varying $\lambda$ quasi-statically from $A$ to $B$.\\
(ii) The system equilibrates with a thermal reservoir at temperature $T$, at fixed $\lambda=B$, reaching state $[B,T]$.\\
(iii) The system is driven {\it isothermally and reversibly} from $[B,T]$ back to the state $[A,T] = (A,S)$.\\
This process is represented schematically as follows:
\be
(A,S) \xrightarrow[adia]{rev} (B,S) \xrightarrow{irrev} [B,T] \xrightarrow[isoth]{rev} [A,T]
\ee
At the end of the first stage, the system is in an equilibrium state $(B,S)$ at temperature $T^*\ne T$, and the second stage consists of placing the system in contact and allowing it to equilibrate with the thermal reservoir (whose temperature is set to that of the system's initial state).
Because the first and third stages are reversible, and no work is performed during the second stage, the work performed over the course of this cyclic process is
\be
W_{cyc}= W_i + W_{iii} = \Delta E_S - \Delta F_T
\ee
By the Kelvin-Planck statement of the second law 
(no process is possible whose sole result is the extraction of energy from a heat bath, and the conversion of all that energy into work~\cite{Thomson1882,Planck1954}) this work must be non-negative, $W_{cyc}\ge 0$, which gives us the desired inequality, Eq.~\ref{eq:strongIneq}.

Eq.~\ref{eq:strongIneq} expresses the general property that adiabats are steeper than isotherms, in phase diagrams where a  thermodynamic force is plotted against the variable conjugate to that force.
This property is most familiarly encountered in the pressure-volume phase diagram of an ideal gas, where it implies that more work is required to compress the gas adiabatically than isothermally, and less work is extracted upon expanding the gas adiabatically than isothermally, assuming a reversible process.

\vskip .5in
\noindent{\it Derivation of $Y = (T^*/T)X$ in the near tail $X\lesssim 0$.}

Let
\be
\rho(X) = \langle \delta \left[ X - E_A(E_f) + E_i \right] \rangle
\ee
denote the distribution of values of $X$, over an ensemble of realizations of an adiabatic process of the sort considered in this paper.
The process is {\it not} assumed to be quasi-static.
Given the initial energy $E_i = H_A(z_0)$, we have
\ba
\label{eq:YofX}
Y &=& E_B(E_i+X) - E_B(E_i) \nonumber \\
\label{eq:YofX2}
&=& \frac{d E_B}{d E_i}(E_i) \cdot X + \frac{1}{2} \frac{d^2 E_B}{d E_i^2}(E_i) \cdot X^2
\ea
neglecting higher-order terms.
Using Eqs.~\ref{eq:EBdef}, \ref{eq:Sigmadef} and \ref{eq:T} we get
\ba
\label{eq:dEdE}
\frac{dE_B}{dE_i}(E_i) &=& \frac{\Sigma_A(E_i)}{\Sigma_B(E_B(E_i))} 
= \frac{T_f^*}{T_i} \\
\frac{d^2 E_B}{dE_i^2}(E_i) &=& \frac{T_f^*}{T_i^2} \, \left( \frac{1}{C_B^*} - \frac{1}{C_A} \right)
\ea
where
\be
T_i(E_i) = T(E_i,A)
\quad , \quad
T_f^*(E_i) = T(E_B(E_i),B)
\ee
are ``microcanonical'' temperatures (Eq.~\ref{eq:T}), and
\be
\begin{aligned}
C_A &= \left[ \frac{\partial T(E,A)}{\partial E} \right]_{E=E_i}^{-1} &= \frac{\partial E(T_i,A)}{\partial T_i} \\
C_B^* &= \left[ \frac{\partial T(E,B)}{\partial E} \right]_{E=E_B(E_i)}^{-1} &= \frac{\partial E(T_f^*,B)}{\partial T_f^*}
\end{aligned}
\ee
We have used the monotonic dependence of temperature on energy to introduce the inverse $E(T,\lambda)$ of the function $T(E,\lambda)$ defined in Eq.~\ref{eq:T}.
The quantities $C_A$ and $C_B^*$ are the heat capacities associated with the microcanonical states $(E_i,A)$ and $(E_B(E_i),B)$.
Eq.~\ref{eq:YofX2} now becomes
\be
\label{eq:Yapprox}
Y = \frac{T_f^*}{T_i} X \cdot (1 + \xi)
\ee
where
\be
\xi = \frac{X}{2T_i}  \left( \frac{1}{C_B^*} - \frac{1}{C_A} \right)
\ee
In the window $X\lesssim 0$ (Eq.~\ref{eq:window}), we have $\vert X\vert/2T_i \sim nk$.
Since the characteristic magnitude of the heat capacity of a macroscopic system is $C \sim N k$, we get
\be
\vert\xi\vert \sim \frac{n}{N} \ll 1
\ee
hence we can neglect the term $\xi$ in Eq.~\ref{eq:Yapprox}, when $X\lesssim 0$.

While the microcanonical temperature $T_i = T(E_i,A)$ differs from one realization of the process to the next, these values are exceedingly narrowly distributed around the canonical temperature $T$ from which the initial conditions are sampled (Eq.~\ref{eq:eqA}).
The values of $T_f^*(E_i) = T(E_B(E_i),B)$ are similarly narrowly distributed around the temperature $T^*$ of the equilibrium state $(B,S)$,
allowing us finally to write 
\be
Y = \frac{T^*}{T} X
\ee
when $X \lesssim 0$.

\section*{References}

\bibliography{refs}

\end{document}